\documentstyle[aps,prl,twocolumn]{revtex}
\begin{document}
\draft

    \title{Inhomogeneous ground state and the coexistence of two
length scales near phase transitions in real solids}

    \author{A. L. Korzhenevskii}
    \address{Electro--Engineering Inst., 197376 St Petersburg, Russia}
    \author{K. Herrmanns, H.--O. Heuer}
    \address{Theoretische Physik III, Ruhr--Universit\"at, 44780 Bochum, 
	Germany}

\date{\today}

\maketitle

\begin{abstract}

Real crystals almost unavoidably contain a finite density of
dislocations. We show that this generic type of long--range correlated
disorder leads to a breakdown of the conventional scenario of critical
behavior and standard renormalization group techniques based on the
existence of a simple, homogeneous ground state. This breakdown is due
to the appearance of an inhomogeneous ground state that changes the
character of the phase transition to that of a percolative phenomenon.
This scenario leads to a natural explanation for the appearance of two
length scales in recent high resolution small-angle scattering
experiments near magnetic and structural phase transitions.

\end{abstract}


\narrowtext

The modern theory of phase transitions (PTs) in pure and weakly
disordered crystals is based on the scaling hypothesis and the
renormalization group (RG). It has been confirmed by a large number of
experiments and numerical simulations.

Therefore, the results of some recent high resolution X--ray and
neutron scattering experiments for various crystals near structural
and magnetic PTs were quite unexpected \cite{Cow96}: in contrast to
the predictions of conventional scaling theory two distinct large
scales were observed. The temperature dependence of the smaller scale
was found to be broadly consistent with the results of the
conventional theory for the correlation scale of thermal
fluctuations. Quite recently explicit experimental evidence for
Verneuil grown samples of SrTiO$_3$ \cite{WZS98} confirmed earlier
speculations that the second, larger scale is connected with the
presence of defects in a very large region near the surface.

All experiments that found two length scales used crystals which were
considered to be of good quality. Nevertheless, surface preparation
creates a significant density of defects in a surface layer with a
thickness up to several $\mu$ \cite{Cow96,WZS98,Wat96,HR82}. Based on
this observation, Altarelli et al.~suggested \cite{ANP95} that most of
the defects are actually dislocation dipoles which induce elastic
strains with a pair correlator $G({\bf r}) \sim |{\bf r}|^{-2}$. They
tried to explain the critical behavior (CB) in the disordered layer as
that of the Weinrib--Halperin (WH) model \cite{WH83}. The coexistence
of two scales was then interpreted as a simple superposition effect,
with scattering from the larger scale in the disordered layer (where
the CB is governed by the WH fixed point) and from the smaller scale
(corresponding to conventional CB) in the bulk of the sample. However,
later experiments using very thin Holmium films found clear evidence
that near the magnetic PT temperature both scales coexist in the same
volume fraction of the sample \cite{Geh96}.

By construction, the conventional RG procedure is not able to explain
the {\em simultaneous} existence of two length scales. Not only its
detailed value and temperature dependence, but even the principal
theoretical basis for the origin of the second length scale appears to
be unclear.

In this Letter we point out the shortcomings of the RG procedure for
the treatment of systems with weak, but long--range (LR) correlated
disorder, the generic situation in crystals with topological defects
like dislocations. We show that the global scenario of the PT differs
from that of models with short--range correlated disorder: in addition
to the thermal correlation length (which does not diverge, but
saturates near $T_c$) even for very small dislocation densities a
second length scale inevitably appears. Its divergence determines the
critical point. In the language of the RG, this scale is connected
with the existence of a temperature dependent inhomogeneous ground
state (GS). This GS is formed by stable nuclei of the ordered phase
that appear even {\em above} the global $T_c$. We compare our results
with experiments for some perovskites and rare earth magnets (REM)
\cite{Cow96}.

We begin with a brief review of the standard RG treatment of the CB of
weakly disordered crystals. The Hamiltonian of the simplest model of
this class, for an $n$--component order parameter $\varphi_i$,
$i=1\dots n$, is given by
\begin{equation}
\label{H}
H = H_{GL} \left\{ \varphi_i \right\} + \sum_i \int d^3 {\bf r} \,
\frac{1}{2} \, \Delta \tau({\bf r}) \varphi_i^2 ({\bf r})
\end{equation}
with the Ginzburg--Landau Hamiltonian $H_{GL}$ for a pure crystal.

The disorder variable $\Delta \tau({\bf r})$ is usually assumed to be
Gaussian distributed, so it is completely characterized by the pair
correlator
\begin{equation}
\label{corrr0}
G({\bf r}) = \langle \Delta \tau({\bf 0}) \Delta \tau({\bf r}) \rangle.
\end{equation}
This can be described by two parameters, a defect strength $\tau_d^2 =
G({\bf 0})$ and a single, finite correlation length $R_0$, so $G({\bf
r}) \to 0$ for $|{\bf r}| \gg R_0$.  For actual RG calculations a
further simplification is usually to ignore the scale $R_0$ and set
\begin{equation}
\label{corrd}
G({\bf r}) = u_0 \, \delta({\bf r}); \ \ \ u_0 = \tau_d^2 \, R_0^d.
\end{equation}
This is selfconsistently justified with the argument that close to
$T_c$ any finite scales like $R_0$ are irrelevant in comparison to the
diverging scale of critical fluctuations, $r_c$ \cite{Litpt}.

However, the RG procedure is a description of the critical
fluctuations around the GS of a system, and its results are valid only
if the GS is determined correctly. Usually, this GS is assumed to be
homogeneous; but in fact it has to be determined using either a Landau
type saddle point equation for a particular disorder configuration
(see below), or the saddle point equation for the disorder averaged
free energy. For uncorrelated Gaussian disorder, the latter can be
written as
\begin{equation}
(\tau - \Delta) \, \varphi_i^{(0)} ({\bf r}) = (u_0 - \lambda_0) \,
\left( \varphi_i^{(0)} ({\bf r}) \right)^3,
\end{equation}
with the pure vertex $\lambda_0$.

If $\lambda_0 \gg u_0$ this equation only has the trivial solution
$\varphi_i^{(0)} ({\bf r}) \equiv 0$ for $\tau>0$. But for stronger
disorder, $u_0 > u_c = \xi \lambda_0$ with $\xi \approx 1$ nontrivial
solutions appear \cite{BPS87}, and we have to take into account the
presence of a nontrivial GS. The standard RG procedure may be used
only if $u_0 \ll \lambda_0$ to describe ``weak'' disorder in the
conventional RG sense \cite{Dot95}. In the opposite case $u_0 > u_c$ a
RG calculation has to account for the presence of an inhomogeneous GS.

This leads to the question what kind of GS exists in ``high quality''
samples of {\em real} solids. The disorder strength $\tau_d$ is small,
so a change of the CB can be expected only when the correlation scale
$R_0$ is so large that the approximation of (\ref{corrr0}) by
(\ref{corrd}) is no longer valid.

Several types of defects lead to LR correlated random
temperature--like disorder. Some special cases, like inhomogeneously
distributed point defects have been discussed before
\cite{IW79,KLS94}. But there is a much more generic class of such
defects: topological defects like dislocations in crystals. Models
with dislocations require a more detailed analysis of many different
situations, with varying types of dislocation ensembles and transition
scenarios.

We begin our analysis of such crystals with dislocations based on the
Landau expansion of the free energy density $f$,
\begin{equation}
\label{f}
f = H_{GL} \left\{ \varphi_i \right\} + q_{ijkl} \,
\epsilon^{(d)}_{kl} ({\bf r}) \ \varphi_{i} \varphi_j.
\end{equation}
The tensor $q_{ijkl}$ describes the coupling \cite{remark} between
the order parameter and the elastic strain introduced by the quenched
disorder, which is given by the sum over contributions from all
dislocations \cite{LL},
\begin{equation}
\label{epsdef}
\epsilon^{(d)}_{kl}({\bf r}) = \sum_{\alpha, i} \frac{b_{\alpha} \,
 s(\phi_i)}{|{\bf r}_{\bot} - {\bf r}_i|},
\end{equation}
where $\phi_i$ is the angle between the component of ${\bf r}_{\bot} -
{\bf r}_i$ perpendicular to the dislocation and the Burgers vector
${\bf b}_{\alpha}$ of a dislocation of type $\alpha$ at ${\bf r}_i$,
and the sum runs over all dislocations. 

To simplify the equations, we will limit the discussion to the case of
a scalar order parameter $\varphi$ and an isotropic elastic medium
\cite{remark}. The GS for a proper RG treatment can be identified
minimizing the free energy (\ref{f}). This leads to the nonlinear
stochastic equation
\begin{equation}
\label{Sch}
- g \ \Delta \varphi + \alpha \ \left(\tau + \frac{q}{\alpha} \
\epsilon^{(d)}_{ii} ({\bf r}) \right) \ \varphi + \lambda \ \varphi^3 +
D \ \varphi^5 = 0.
\end{equation}
The simplified coupling is $q/\alpha \sim (K/T_0) \, \left( \partial
T_c / \partial p \right)$ with the elastic modulus $K$, the Curie
temperature $T_0$ of the undisturbed system and pressure $p$. The
inclusion of a $\varphi^5$--term allows us to discuss the weakly first
order structural PT in perovskites.

The statistical properties of the solutions of (\ref{Sch}) are
determined by the random elastic strain $\epsilon =
\epsilon^{(d)}_{ii}$. Obviously the coefficient of the linear term in
(\ref{Sch}) can become negative and allow a nonzero solution of
(\ref{Sch}), either due to the large contribution to $\epsilon ({\bf
r})$ of a single dislocation close to ${\bf r}$, or because the
contributions of several dislocations happen to add up to a large
(negative) value of $\epsilon$ at ${\bf r}$.

For randomly distributed dislocations loops with a radius of curvature
$L$ much larger than the average distance $r_d$ between loops
\cite{loops} the probability distribution function for $\epsilon$ in
(\ref{Sch}) is a Levy distribution \cite{remarkLevy}: ${\cal
P}(\epsilon) = {\cal P}^{(L)}_{\mu=2} (\epsilon)$, in particular
${\cal P}(\epsilon) \sim \epsilon^{-3}$ for $\epsilon \gg 1$. The
variance of this distribution is given by
\begin{equation}
\label{vare}
\mbox{var} \ \epsilon \approx \left( \epsilon (r_d) \right)^2 \ \ln
(L/r_d) \equiv \left( \frac{\alpha}{q} \right)^2 \tau_1^2
\end{equation}
with $\epsilon (r_d) \approx b/2 \pi \, r_d$ and the corresponding
temperature scale $\tau_1$.

The pair correlation function for this distribution for $r \le L$ is
$G(r) = \langle \epsilon (0) \, \epsilon ({\bf r}) \rangle \sim n_d \,
\ln (L/r) $ with the dislocation density $n_d = r_d^{-2}$. It is
determined by the contribution of many dislocations on scales $r_{min}
< r < L$, while for distances $r < r_{min} = r_d \, (\ln (L/r_d))^{-
1/2}$ the random value of $\epsilon ({\bf r})$ is dominated by the
contribution of the single nearest dislocation.

As the elastic strains are strongest in the immediate vicinity of the
dislocations, nonzero solutions of (\ref{Sch}) may first appear there
as the temperature is lowered from high values. In a first
approximation, we can limit (\ref{Sch}) to include only the strain
induced by the closest dislocation $i$ near ${\bf r}$:
\begin{equation}
\label{SchSDA}
- g \ \Delta \varphi + \alpha \ \left(\tau + \frac{q}{\alpha} \
\frac{b \, s(\phi)}{|{\bf r}_{\bot} - {\bf r}_i|} \right) \
\varphi + \lambda \ \varphi^3 + D \ \varphi^5 = 0.
\end{equation}

The temperature $T_n$ at which nonzero solutions of this nonlinear,
deterministic equation first appear can be found rigorously
\cite{KHS96} by an analysis of the linear part of this equation. The
main results of this analysis are as follows. The reduced temperature
$\tau_n = (T_n - T_0)/T_0$ and the characteristic scale of ordered
nuclei $R(\tau_n)$ are,
\begin{eqnarray}
\tau_n & \approx & \frac{1}{\alpha g} \, (q \, b)^2 + \Delta \tau_h,
\\
\label{R(tn)}
R(\tau_n) & \approx & \frac{g} {q \, b} \approx \sqrt{ \frac{g} {\alpha
\tau_n} } = r_c(\tau_n). 
\end{eqnarray}
Here $\Delta \tau_h \ll \tau_n$ describes the temperature range of the
hysteresis in a crystal without dislocations and a weakly first order
PT like the perovskites that display the two length scales
\cite{BC80}.  The second expression for $R(\tau_n)$ reflects the fact
that nonzero solutions appear when the energy gain from the attractive
potential balances the ``kinetic'' energy due to the gradient term in
the free energy.

The above reasoning depends on two assumptions: the possibility to
include only the simplest gradient term in (\ref{Sch}) and
(\ref{SchSDA}), and the validity of the single defect approximation
(\ref{SchSDA}).

Obviously, the former requirement holds if $R(\tau_n) \gg a$, where
$a$ is the lattice constant. The coefficient $\alpha$ can be written
as $\alpha = T_0 / C$ with the material's Curie constant $C$. In
perovskites, for displacive PTs \cite{BC80} $C \approx 10^4$ -- $10^5
K$, so $\alpha = O(10^{-2}$ -- $10^{-3})$ and, indeed, $R(\tau_n)
\approx a/\alpha \gg a$. The scattering experiments for SrTiO$_3$ also
show \cite{McM90} that in the relevant temperature range $r_c \gg a$,
so the use of (\ref{Sch}) and (\ref{SchSDA}) is justified.

In REM, $C \ll T_0$ \cite{Shi81}, so $\alpha$ is large and (\ref{Sch})
and (\ref{SchSDA}) cannot be used at all. However, in both cases below
$T_n$ the radius $R(\tau)$ grows much larger than $r_c$, so one can
use the estimation 
\begin{equation}
\label{R(T)gen}
R(\tau) \approx \frac{K}{T_0} \, \left( \partial T_c / \partial p
\right) \, \frac{b}{2 \pi \ \tau}
\end{equation}
in the temperature range where $R(\tau) \gg r_c(\tau)$. Obviously the
``divergence'' of the scale $R(\tau)$ is described by an exponent
$\nu=1$. The prefactor in (\ref{R(T)gen}) has very similar numerical
values in perovskites and REM, which leads to scales of a few 10$^3$
\AA \ in both classes \cite{tobe}. In the temperature range where
(\ref{R(T)gen}) is valid, the integrated intensity of scattering from
the nuclei of the ordered phase is $I(\tau) \sim n_d R^2(\tau) \sim
\tau^{-2}$.

As the temperature is reduced the size and the number of nuclei
grow. To determine the global character of the PT one must then
investigate the temperature dependence of the disorder averaged GS
correlator $g({\bf r}) = \langle \varphi^{(0)} ({\bf 0}) \
\varphi^{(0)} ({\bf r}) \rangle$ for $r \to \infty$. For this purpose
it is crucial to analyze the state of ``bridges'' with $\tau({\bf r})
< 0$ which link two nuclei that have appeared at some higher
temperature \cite{fn5}.

It is easy to show \cite{KHH98,tobe} that if on such a bridge
$\tau({\bf r}) < - \tau_2 \equiv g/\alpha \, r^2_{min}$ the
correlation radius of local thermal fluctations is $r_c({\bf r}) \ll
r_{min}$. In such bridges the energy of a domain wall (DW) is much
larger than $T_c$, so no DWs will appear. If $- \tau_2 < \tau({\bf r})
< 0$ then $r_c({\bf r}) \ge r_{min}$ and DWs are easy to form. The
probability for this is $Prob(r_c({\bf r}) \ge r_{min}) \approx
(\tau_2 / \tau_1)^{1/2} \equiv z$. The sign of the order parameter in
the nuclei connected by a bridge is the same with a probability $1-z
\approx 1$ for $z\ll 1$. Of course, this sign will fluctuate in time
as long as overlapping nuclei form finite clusters. 

As the temperature is decreased, these clusters will grow, and at
$T_c$ an infinite percolating cluster will appear. On this cluster the
sign of the order parameter cannot change anymore --- this is the
spontaneous symmetry breaking that defines the phase transition in the
system. Therefore the global PT scenario is a percolative one. The
condition $z \ll 1$ can be rewritten as $R(\tau_n) \ll r_{min}$ using
(\ref{R(tn)}). This turns out to be equivalent to the condition under
which the first nucleation of the ordered phase near dislocations can
be described in the single defect approximation (\ref{SchSDA}).

The smaller length scale seen in experiment corresponds to the
fluctuations $\tilde \varphi ({\bf r}) = \varphi ({\bf r}) -
\varphi^{(0)} ({\bf r})$ around the inhomogeneous, temperature
dependent GS $\varphi^{(0)} ({\bf r})$ determined by (\ref{Sch})
\cite{fn}. Assuming mean field exponents for simplicity, the width of
the corresponding ``broad component'' in momentum space can be
estimated as
\begin{eqnarray}
(\Delta q)^2 = & \langle r_c^{-2} \rangle = A_+ \int_0^{\infty}
\tau' \, {\cal P} (\epsilon = \frac{\alpha}{q} (\tau'-\tau)) \, d
\tau' 
\nonumber \\  
& + A_- \int_{-\infty}^0 |\tau'| \, {\cal P} (\epsilon =
\frac{\alpha}{q} (\tau'-\tau)) \, d\tau'
\end{eqnarray}
with $\tau = (T-T_0)/T_0$. This implies that the line width remains
finite at all temperatures, with a minimal value $\Delta q_{min}
\approx r_c^{-1}(\tau_1)$. This is reached (in the mean field
approximation, $A_-/A_+ = 2$) at $p = 1/3$. Because in three
dimensions the percolation concentration $p_c < 1/3$, this minimum
lies slightly below the temperature $T_c$ where the length scale
connected with the narrow component of the scattering diverges, as
observed in experiments \cite{HSGM94}.

If $R(\tau_n) > r_{min}$, neither the single defect approximation nor
the simple percolative picture of the PT hold, as the inhomogeneous GS
$\varphi^{(0)}({\bf r})$ and thermal fluctuations both play an equally
important role. In this case a completely new generalized RG scheme is
needed.

To summarize, we have analyzed the CB of solids with LR correlated
quenched disorder taking into account the appearance of an
inhomogeneous, temperature dependent GS. In real crystals the
appearance of such a GS is a generic phenomenon due to the almost
unevitable presence of dislocations. The appearance of this GS is
important even for a qualitative discussion; in particular it leads to
the breakdown of the conventional RG procedure.

We have shown that if the condition $R(\tau_n) \ll r_{min}$ holds the
global PT scenario consists of two stages: at $\tau = \tau_n$ nuclei
of the low-temperature ordered phase appear next to individual
dislocation lines. At lower $\tau$ the nuclei grow and begin to
overlap; the typical size of magnetically ordered domains is described
by percolation theory. It is especially remarkable that this PT
scenario becomes more likely when $n_d$ is reduced, because $r_{min}
\propto n_d^{-1/2}$. So generally speaking the two length scale
phenomenon should be observable in scattering experiments in many
crystals. Of course a high resolution setup with sufficient
sensibility as used in refs.~\onlinecite{Wat96,HSGM94,Geh96} is
necessary, especially for {\em better} samples with small $n_d$,
because the volume occupied by nuclei of the ordered phase becomes
noticable at temperatures of order $\tau_1 \sim n_d^{1/2}$ (using
(\ref{vare}) and (\ref{R(T)gen}) one can write $\tau_1 = (n_d \,
b^2)^{1/2}$) for most structural PTs and the PT in REM.

The universality of the percolation PT scenario due to LR correlated
disorder can be illustrated by the the recent observation \cite{Gri97}
of two scales in the neutron scattering data for the Invar alloy
Fe$_{1-x}$Ni$_x$ which is near a morphology boundary for $x\approx
0.3$. Contrary to earlier speculation \cite{Litpt} the description of
the CB of {\em real} crystals very close to $T_c$ must account for LR
correlated disorder in addition to the uncorrelated disorder covered
by the conventional theory. In fact, the asymptotic CB of ``high
quality'' samples of perovskites and REM is the same as that of a
``dirty'' alloy with LR chemical disorder \cite{RSW97}.

This work was supported by SFB 237 (Disorder and Large Fluctuations)
and by RFFI grants N16893 and 16958.


\begin{references}
\bibitem{remark} More complicated coupling terms like $q_{xyxy} \,
(|\varphi_x|^2 + |\varphi_y|^2) \, \epsilon^{(d)}_{xy}$ or gradient
terms like $g_{ijkl} \, \partial_i \varphi_j \, \partial_k \varphi_l$
are important for a quantitative description of real materials, but
omitted in the present outline of the general idea.
\bibitem{loops} Small dislocation loops can be discussed using the
same general line of reasoning.
\bibitem{fn5} In fact, the topology of such bridges corresponds to
that of multiply connected red sites of a percolation cluster. The
appearance and temperature evolution of domain walls (DWs) on such
bridges for the simple model (\ref{corrr0}) have been discussed in
ref.~\onlinecite{KHH98}. A detailed analysis for crystals with
dislocations will be published elsewhere.
\end{references}
\end{document}